\documentclass[14pt]{article}
\usepackage{amssymb}
\usepackage{amsmath}
\usepackage{graphicx}
\usepackage[pagebackref=false,colorlinks=,linkcolor=blue,citecolor=blue]{hyperref}
\usepackage{subcaption}
\usepackage{bm}
\newcommand{\mathsym}[1]{{}}

\usepackage{graphicx}
\usepackage{rotating}
\usepackage{color}

\newcommand{\bra}{\begin{array}}
	\newcommand{\era}{\end{array}}
\newcommand{\beq}{\begin{equation}}
\newcommand{\eeq}{\end{equation}}
\newcommand{\beqar}{\begin{eqnarray}}
\newcommand{\eeqar}{\end{eqnarray}}

\newcommand{\be}{\begin{equation}}
\newcommand{\ee}{\end{equation}}
\newcommand{\bea}{\begin{eqnarray}}
\newcommand{\eea}{\end{eqnarray}}
\newcommand{\bd}{\begin{displaymath}}
\newcommand{\ed}{\end{displaymath}}

\begin{document}
	\baselineskip=18 pt
	\begin{center}
		{\large {\bf
				Energy corrections of the two-dimensional Dunkl-harmonic oscillator in non-commutative phase-space}}
	\end{center}\color{black}
	\vspace{.5cm}
	
	\begin{center}
	{\bf S. Hassanabadi  $^{1,}$}\footnote{ s.hassanabadi@yahoo.com},\, {\bf P. Sedaghatnia  $^{2,}$}\footnote{pa.sedaghatnia@gmail.com},\,
		{\bf W. S. Chung $^{3,}$}\footnote{mimip44@naver.com},\,
		{\bf B. C. L${\bf\ddot{u}}$tf${\bf\ddot{u}}$o${\bf\breve{g}}$lu $^{1,4,}$}\footnote{bclutfuoglu@akdeniz.edu.tr(Corresponding author)}\\ 
		{\bf  J. K${\bf \breve{r}}$i${\bf \breve{z}}$ $^{1,}$}\footnote{jan.kriz@uhk.cz} and {\bf H.  Hassanabadi $^{1,2}$}\footnote{h.hasanabadi@shahroodut.ac.ir}
		    \\
		    	\small \textit {\it $^{1}$ Department of Physics, University of Hradec Kr$\acute{a}$lov$\acute{e}$,
		    	Rokitansk$\acute{e}$ho 62, 500 03 Hradec Kr$\acute{a}$lov$\acute{e}$, Czechia.}\\
		    \small \textit {\it $^{2}$ Faculty of Physics, Shahrood University of Technology, Shahrood, Iran \\ P. O. Box : 3619995161-316.}
		  \\
		\small \textit {\it $^{3}$
		Department of Physics and Research Institute of Natural Science, College of Natural Science, \\Gyeongsang National University, Jinju 660-701, Korea.}
	\\
	\small \textit {\it $^{4}$ Department of Physics, Akdeniz University, Campus 07058, Antalya, Turkey.} 
	\end{center}
\vspace{.5cm}
\begin{abstract}
In this paper, we examine the harmonic oscillator problem in non-commutative phase space (NCPS) by using the Dunkl derivative instead of the habitual one. After defining the Hamilton operator,  we use the perturbation method to derive the binding energy eigenvalues. We find eigenfunctions that correspond to these eigenvalues in terms of the associated Laguerre polynomials. We observe that the Dunkl-Harmonic Oscillator (DHO) in the NCPS differs from the ordinary one in the context of providing additional information on the even and odd parities. Therefore, we conclude that working with the Dunkl operator could be more appropriate because of its rich content. 
\end{abstract}



\section {Introduction}

In the last three decades, the study of physical problems with the concept of non-commutativity of the coordinates  based on various motivations has received a great deal of attention \cite{4,1,2,3,3a,3b, 5,6,heddar,nc1}. In Ref. \cite{2}, the authors presented one of the first formulations of examining the non-commutativity of the coordinate space within the usual non-relativistic quantum mechanics. Since the formulation of NCPS is complex, most of the quantum mechanical problems, for example, the central potential problems, can be solved within perturbation methods \cite{3a,3b}. In Refs. \cite{5,6,heddar}, the authors investigated the Klein-Gordon and Pauli-Dirac oscillator dynamics in NCPS. Recently in a very interesting paper authors considered a charged harmonic oscillator in NCPS and showed that if a uniform magnetic field exists, then a minimum length and a minimum momentum uncertainty may arise \cite{nc1}.

Lately, we observe an increasing interest in studies dealing with the Dunkl derivative. This is not surprising, since the Dunkl derivative operator has a rich structure consisting of a combination of differential and discrete parts with the reflection operator term \cite{Dunkl1,Dunkl2}. In fact, the history of this deformation dates back to the middle of the last century. In 1951, Yang employed the reflection operator, which was first defined by Wigner one year ago \cite{Wigner1}, to solve one-dimensional quantum harmonic oscillator \cite{Yang1}. In 2013, in a series of two papers, Genest et al. considered an isotropic DHO in the plane by displacing the Dunkl derivative with the partial derivative and discussed the superintegrability of the model  \cite{x23, x32}. Then, they 
revisited the problem in three spatial dimensions \cite{x33}. Two years later, they examined the non-relativistic Dunkl-Coulomb problem in the same context \cite{x26}. In 2019, Mota et al. used this approach in the relativistic regime and obtained an exact solution of the Dunkl-Dirac oscillator which is under the influence of an external magnetic field \cite{mota 1}.
Last year, they solved the Dunkl-Klein-Gordon equation in two dimensions with the Coulomb potential \cite{mota 2}. Very recently, the Dunkl-Klein-Gordon oscillator solutions are found in two and three dimensions, respectively in \cite{mota 3, hamil}. 

We believe that the Dunkl derivative formalism in a  fuzzy space could provide very interesting results in theoretical physics.  With this motivation, we consider a two-dimensional DHO in noncommutative space and intend to derive the energy eigenvalues and their corresponding eigenfunctions within perturbation methods.   We organize the manuscript as follows:  In section \ref{sec1}, we construct the two dimensional Dunkl-Hamiltonian operator of the harmonic oscillator in the NCPS. Then, in section \ref{sec2}, we employ perturbation techniques to obtain eigenenergies and their corresponding eigenstates.
In the final section, we conclude the manuscript.

\section{Two dimensional non-commutative Dunkl-harmonic oscillator}\label{sec1}

The time-independent perturbation theory is based on the assumption in which the Hamiltonian of the system can be divided into two parts, such that an exact solution of one of them can be obtained and the other part can be approximated with the help of these solutions. Therefore, we have to derive the Hamiltonian of the system we are considering. To this end, at first we need to express the non-commutative Hamiltonian, and then, extend it by replacing the partial  derivatives with the  Dunkl ones. Accordingly, we start by writing the non-commutative harmonic oscillator Hamiltonian in two dimensions
\begin{equation}\label{1}
H^{NC}=\frac{1}{2m}\bigg[(P_{x}^{NC})^{2}+(P_{y}^{NC})^{2}\bigg]+\frac{1}{2}m\omega^{2}\bigg[(x^{NC})^{2}+(y^{NC})^{2}\bigg],
\end{equation}
where the non-commutative momentum and position operators are defined as \cite{nc1}
\begin{subequations}
\begin{eqnarray}
P_{x}^{NC}&=&P_{x}+\frac{\eta}{2}y, \label{2a} \\
P_{y}^{NC}&=&P_{y}-\frac{\eta}{2}x, \label{2b}\\
x^{NC}&=&x-\frac{\theta}{2}P_{y}, \label{2c} \\
y^{NC}&=&y+\frac{\theta}{2}P_{x}. \label{2d}
\end{eqnarray}	
\end{subequations}
Here, $\eta$ and $\theta$ are the NCPS parameters. Then, we substitute these operators in Eq. \eqref{1}, and express the non-commutative Hamilton operator in the form of
\begin{eqnarray}\label{3}
H^{NC}=\bigg[\frac{(1+m^{2} \omega^{2}\frac{\theta^{2}}{4})}{2m}\bigg](p_{x}^{2}+p_{y}^{2})+\bigg[\frac{(\frac{\eta^{2}}{4}+m^{2} \omega^{2})}{2m}\bigg]({x}^{2}+{y}^{2})+\bigg[\frac{(\eta+m^{2} \omega^{2}\theta)}{2m}\bigg](yp_{x}-xp_{y}).
\end{eqnarray}
Next, we change the partial derivatives with the Dunkl ones \cite{x23}. It is worth noting that partial derivatives only appear in the momentum operators, $p_{x}$ and $p_{y}$. Therefore, we consider
\begin{eqnarray}
p_{x}\longrightarrow p_{x}^{\mu_{1}}=\frac{\hbar}{i} D_{x}^{\mu_{1}}&=&\frac{\hbar}{i}\bigg[\frac{\partial}{\partial x}+\frac{\mu_{1}}{x}(1-R_{1})\bigg], \label{4} \\
p_{y}\longrightarrow p_{y}^{\mu_{2}}=\frac{\hbar}{i} D_{y}^{\mu_{2}}&=&\frac{\hbar}{i}\bigg[\frac{\partial}{\partial y}+\frac{\mu_{2}}{y}(1-R_{2})\bigg], \label{5}
\end{eqnarray}
where the Wigner parameters, $\mu_1$ and $\mu_2$, are positive two constants \cite{x26}. Besides, the reflection operators, $R_{1}$ and $R_{2}$, satisfy
\begin{eqnarray}\label{6}
R_{1}f(x,y)=f(-x,y)\quad,\quad R_{2}f(x,y)=f(x,-y)\quad,\quad R_{1}^{2}=1\quad,\quad R_{1} x=-xR_{1}\quad,\quad R_{2} y=-yR_{2}.
\end{eqnarray}
By using the Dunkl-momentum operators defined in equations \eqref{4} and \eqref{5}, we construct all necessary operators of the Hamiltonian, such as
\begin{eqnarray}
p_{x}^{2}+p_{y}^{2}\longrightarrow -\hbar^{2}\bigg[\frac{\partial^{2}}{\partial x^{2}}+\frac{2\mu_{1}}{x}\frac{\partial}{\partial x}-\frac{\mu_{1}}{x^{2}}(1-R_{1})+\frac{\partial ^{2}}{\partial y^{2}}+\frac{2\mu_{2}}{y}\frac{\partial}{\partial y}-\frac{\mu_{2}}{y^{2}}(1-R_{2})\bigg], \label{8}
\end{eqnarray}
and the $z-$component of the Dunkl-angular momentum operator
\begin{eqnarray}\label{7}
yp_{x}-xp_{y}\longrightarrow \frac{\hbar}{i}\bigg[ y\frac{\partial}{\partial x}-x \frac{\partial}{\partial y}+\frac{y}{x}\mu_{1}(1-R_{1})-\frac{x}{y}\mu_{2}(1-R_{2})\bigg].
\end{eqnarray}
After performing some simple algebra, we express the non-commutative DHO Hamiltonian
as 
\begin{eqnarray}\label{9}
H_{Dunkl}=\mathcal{G}_{1}H_{0}(x,y)+\mathcal{G}_{2}H_{1}(x,y)
\end{eqnarray}
where
\begin{eqnarray}\label{10}
&&H_{0}
(x,y)=\frac{\partial^{2}}{\partial x^{2}}+\frac{2\mu_{1}}{x}\frac{\partial}{\partial x}-\frac{\mu_{1}}{x^{2}}(1-R_{1})+\frac{\partial ^{2}}{\partial y^{2}}+\frac{2\mu_{2}}{y}\frac{\partial}{\partial y}-\frac{\mu_{2}}{y^{2}}(1-R_{2})+\mathcal{G}_{3}(x^{2}+y^{2})+\mathcal{G}_{4}L_{z},\\
&&H_{1}(x,y)=\frac{y}{x}\mu_{1}(1-R_{1})-\frac{x}{y}\mu_{2}(1-R_{2}),\\
&&\mathcal{G}_{1}=\frac{-\hbar^{2}(1+m^{2}\omega^{2}\frac{\theta^{2}}{4})}{2m}\quad,\quad \mathcal{G}_{2}=\frac{\hbar(\eta+m^{2}\omega^{2}\theta)}{2i m} \quad,\quad \mathcal{G}_{3}=\frac{-(\frac{\eta^{2}}{4}+m^{2}\omega^{2})}{\hbar^{2}(1+m^{2}\omega^{2}\frac{\theta^{2}}{4})},\quad\mathcal{G}_{4}=-\frac{\quad\mathcal{G}_{2}}{\quad\mathcal{G}_{1}}.
\end{eqnarray}
Obviously, for $\mu_{1}=\mu_{2}=\theta=\eta= 0$, we return to the ordinary case.

\section{First-order perturbation }\label{sec2}

In this section, we intend to derive the binding energy of the two dimensional DHO via the first-order perturbation theory. In order to state the impact of noncommutativity, first we revisit the well-known commutative case, where the Dunkl-Hamiltonian satisfies 
\begin{eqnarray}\label{13}
H_{0}(x,y)\psi_{n_{x},n_{y}}^{R_{1},R_{2}}(x,y)=E_{n_{x},n_{y}}^{(0)}\psi_{n_{x},n_{y}}^{R_{1},R_{2}}(x,y).
\end{eqnarray}
We assume that the wavefunction can be separable to
\begin{eqnarray}\label{14}
\psi_{n_{x},n_{y}}^{(R_{1},R_{2})}(x,y)=\psi_{n_{x}}^{R_{1}}(x)\psi_{n_{y}}^{R_{2}}(y).
\end{eqnarray}
Then, we substitute it into equation \eqref{13}. Straightforwardly, we obtain the following eigenenergies
\begin{eqnarray}\label{16}
E^{(R_{1},R_{2})}_{n_{x},n_{y}}=\mathcal{G}_{1} \mathcal{G}_{4} L_{z}-\mathcal{G}_{1} \sqrt{\mathcal{G}_{3}} \bigg(4 (n_{x}+n_{y}+1)+\mathcal{K}_{1}+\mathcal{K}_{2}\bigg),
\end{eqnarray}
and eigenstates
\begin{eqnarray}\label{15}
&&\psi^{R_{1}}_{n_{x}}(x)=e^{-\frac{1}{2} \sqrt{\mathcal{G}_{3}} x^2} x^{\frac{1}{2} \left(-2 \mu_{1}+\mathcal{K}_{1}+1\right)} L_{n_x}^{\frac{1}{2} \mathcal{K}_{1}}\left(\sqrt{\mathcal{G}_{3}} x^2\right),\\
&&\psi^{R_{2}}_{n_{y}}(y)=e^{-\frac{1}{2} \sqrt{\mathcal{G}_{3}} y^2} y^{\frac{1}{2} \left(-2 \mu_{2}+\mathcal{K}_{2}+1\right)} L_{n_y}^{\frac{1}{2} \mathcal{K}_{2}}\left(\sqrt{\mathcal{G}_{3}} y^2\right),
\end{eqnarray}
where $\mathcal{K}_{i}=\sqrt{4 \mu_{i}^{2}-4\mu_{i} R_{i}+1}$, for $i=1,2$. Here, $L_{n}^{\alpha}(u)$ denotes the associated Laguerre
polynomial. According to the  first-order perturbation theory, we
evaluate the unperturbed energy with the help of equation \eqref{9} as follows:
\begin{eqnarray}\label{19}
\textbf{$(R_{1},R_{2})$}:\quad\quad\quad E^{(0)}_{n_{x},n_{y}}=\langle n_{x},n_{y}|\mathcal{G}_{1} H_{0}|n_{x},n_{y}\rangle=\mathcal{G}_{1}^{2}\bigg[ \mathcal{G}_{4} m \hbar- \sqrt{\mathcal{G}_{3}} \bigg(4 (n_{x}+n_{y}+1)+\mathcal{K}_{1}+\mathcal{K}_{2}\bigg)\bigg]
\end{eqnarray}
where $L_{z}=m\hbar$. Due to the presence of the parity operators $(R_{1},R_{2})$ in equation \eqref{19}, it is possible to obtain degenerate energy eigenvalues for different parity modes.
\begin{eqnarray}
\textbf{$(odd-odd)$}:\quad\quad &&E_{0,0}^{(0)}=\mathcal{G}_{1}^{2}\bigg(\mathcal{G}_{4}m \hbar -2 \sqrt{\mathcal{G}_{3}} (\text{$\mu $}_{1}+\text{$\mu $}_{2}+3)\bigg).\\
&&E_{0,1}^{(0)}=\mathcal{G}_{1}^{2}\bigg(\mathcal{G}_{4}m \hbar -2 \sqrt{\mathcal{G}_{3}} (\text{$\mu $}_{1}+\text{$\mu $}_{2}+5)\bigg)=E_{1,0}^{(0)}.\\
&&E_{1,1}^{(0)}=\mathcal{G}_{1}^{2}\bigg(\mathcal{G}_{4}m \hbar -2 \sqrt{\mathcal{G}_{3}} (\text{$\mu $}_{1}+\text{$\mu $}_{2}+7)\bigg).\\
&&
\nonumber\\
\textbf{$(even-even)$}:\quad\quad &&E_{0,0}^{(0)}=\mathcal{G}_{1}^{2}\bigg(\mathcal{G}_{4}m \hbar -2 \sqrt{\mathcal{G}_{3}} (\text{$\mu $}_{1}+\text{$\mu $}_{2}+1)\bigg).\\
&&E_{0,1}^{(0)}=\mathcal{G}_{1}^{2}\bigg(\mathcal{G}_{4}m \hbar -2 \sqrt{\mathcal{G}_{3}} (\text{$\mu $}_{1}+\text{$\mu $}_{2}+3)\bigg)=E_{1,0}^{(0)}.\\
&&E_{1,1}^{(0)}=\mathcal{G}_{1}^{2}\bigg(\mathcal{G}_{4}m \hbar -2 \sqrt{\mathcal{G}_{3}} (\text{$\mu $}_{1}+\text{$\mu $}_{2}+5)\bigg).\\
&&
\nonumber\\
\textbf{$(even-odd)$$=(odd-even)$}:\quad\quad &&E_{0,0}^{(0)}=\mathcal{G}_{1}^{2}\bigg(\mathcal{G}_{4}m \hbar -2 \sqrt{\mathcal{G}_{3}} (\text{$\mu $}_{1}+\text{$\mu $}_{2}+2)\bigg).\\
&&E_{0,1}^{(0)}=\mathcal{G}_{1}^{2}\bigg(\mathcal{G}_{4}m \hbar -2 \sqrt{\mathcal{G}_{3}} (\text{$\mu $}_{1}+\text{$\mu $}_{2}+4)\bigg)=E_{1,0}^{(0)}.\\
&&E_{1,1}^{(0)}=\mathcal{G}_{1}^{2}\bigg(\mathcal{G}_{4}m \hbar -2 \sqrt{\mathcal{G}_{3}} (\text{$\mu $}_{1}+\text{$\mu $}_{2}+6)\bigg).
\end{eqnarray}
Essentially, in general we can write
\begin{eqnarray}
&&E^{(0)}_{(n_{x},n_{y};even,even)}=E^{(0)}_{(n_{y},n_{x};even,even)},\\
&&E^{(0)}_{(n_{x},n_{y};odd,odd)}=E^{(0)}_{(n_{y},n_{x};odd,odd)},\\
&&E^{(0)}_{(n_{x},n_{y};odd,even)}= E^{(0)}_{(n_{y},n_{x};odd,even)}.
\end{eqnarray}

If we consider weak non commutativity, then we can determine energy eigenvalue corrections to the binding energy.  According to the first-order perturbation theory, these corrections have to be calculated with the help of equation (11) via
\begin{eqnarray}
\textbf{$(R_{1},R_{2})$}:\quad\quad\quad\Delta E^{(1)}_{n_{x},n_{y}}=\mathcal{G}_{2}\langle n_{x},n_{y}|\frac{y}{x}\mu_{1}(1-R_{1})-\frac{x}{y}\mu_{2}(1-R_{2})|n_{x},n_{y}\rangle.
\end{eqnarray}
We find
\begin{eqnarray}
\textbf{$(odd-odd)$}:\quad&&\Delta E_{0,0}^{(1)}=\frac{\alpha_{1}}{4}\bigg(\text{$\mu $}_{1}-2 \text{$\mu $}_{2}\bigg).\\
&&\Delta E_{0,1}^{(1)}=\frac{\alpha_{1}}{8}\bigg(\text{$\mu $}_{1} (4 (\text{$\mu $}_{2}-1) \text{$\mu $}_{2}+9)-\text{$\mu $}_{2} (4 \text{$\mu $}_{2} (\text{$\mu $}_{2}+1)+5)\bigg).
\\
&&\Delta E_{1,0}^{(1)}=\frac{\alpha_{1}}{16}\bigg(\mu_{1} (4 (\text{$\mu $}_{2}-1) \text{$\mu $}_{2}+9)-2 \text{$\mu $}_{2} (4  (\text{$\mu $}_{2}+1)\text{$\mu $}_{2}+5)\bigg).\\
&&\Delta E_{1,1}^{(1)}=\frac{\alpha_{1}}{64}\bigg(4 \text{$\mu $}_{1}^3 (4 (\text{$\mu $}_{2}-1) \text{$\mu $}_{2}+9)-4 \text{$\mu $}_{1}^2 (2 \text{$\mu $}_{2}-1) (4 \text{$\mu $}_{2} (\text{$\mu $}_{2}+1)+9)\nonumber\\
&&+\text{$\mu $}_{1} (4 \text{$\mu $}_{2} (\text{$\mu $}_{2}+1) (8 \text{$\mu $}_{2}+5)+45)-18 \text{$\mu $}_{2} (4 \text{$\mu $}_{2} (\text{$\mu $}_{2}+1)+5)\bigg).\\
&&
\nonumber\\
\textbf{$(even-even)$}:\quad&&\Delta E_{n_{x},n_{y}}^{(1)}=0.
\end{eqnarray}
\begin{eqnarray}
\textbf{$(even-odd)$}:\quad&&\Delta E_{0,0}^{(1)}=-\frac{\alpha_{2}\mu_{2}}{2}.\\
&&\Delta E_{0,1}^{(1)}=-\frac{\alpha_{2}\mu_{2}}{8}\bigg(4  (\text{$\mu $}_{2}+1)\text{$\mu $}_{2}+5\bigg).\\
&&\Delta E_{1,0}^{(1)}=-\frac{\alpha_{2}\mu_{2}}{8}\bigg(4 (\text{$\mu $}_{1}-1) \text{$\mu $}_{1}+5\bigg).\\
&&\Delta E_{1,1}^{(1)}=-\frac{\alpha_{2}\mu_{2}}{32}\bigg(4 (\text{$\mu $}_{1}-1) \text{$\mu $}_{1}+5\bigg) \bigg(4  (\text{$\mu $}_{2}+1)\text{$\mu $}_{2}+5\bigg).
\end{eqnarray}
\begin{eqnarray}
\textbf{$(odd-even)$}:\quad&&\Delta E_{0,0}^{(1)}=\frac{\alpha_{2}\mu_{1}}{2}.\\
&&\Delta E_{0,1}^{(1)}=\frac{\alpha_{2}\mu_{1}}{8}\bigg(4 (\text{$\mu $}_{2}-1) \text{$\mu $}_{2}+5\bigg).\\
&&\Delta E_{1,0}^{(1)}=\frac{\alpha_{2}\mu_{1}}{8}\bigg(4 (\text{$\mu $}_{1}+1) \text{$\mu $}_{1}+5\bigg).\\
&&\Delta E_{1,1}^{(1)}=\frac{\alpha_{2}\mu_{1}}{32}\bigg(4 (\text{$\mu $}_{1}+1) \text{$\mu $}_{1}+5\bigg) \bigg(4  (\text{$\mu $}_{2}-1)\text{$\mu $}_{2}+5\bigg).
\end{eqnarray}
Here, $\alpha_{1}=\frac{\mathcal{G}_{2}}{ \mathcal{G}_{3}^{3/2}}$, and $\alpha_{2}=\frac{\mathcal{G}_{2}}{ \mathcal{G}_{3}}$.  These results hold for the situations where $\eta<2 m \omega$, $\theta<\frac{\hbar}{m \omega}$, and $\mu_{i}>-\frac{1}{2}$.

\section{Conclusion}
In this manuscript, we examine two dimensional DHO problem in a non-commutative phase space. In the context of time-independent perturbation theory, we derive eigenenergies and the corresponding eigenstates in terms of associated  Laguerre polynomials.
We show that considering Dunkl operator leads to different results than its ordinary state according to 
the even and odd parities. Our results also points out the  degeneracy in energy eigenvalues. 

\section*{Acknowledgments}
This work is supported by the Internal  Project,  [2022/2218],  of  Excellent  Research  of  the  Faculty  of  Science  of  University.

\section*{Data Availability Statements}
The authors declare that the data supporting the findings of this study are available within the article.

\end{document}